# Capacitive imaging using fused amplitude and phase information for improved defect detection

Silvio Amato[1], David. A. Hutchins[1], Xiaokang Yin[2], Marco Ricci[3], and Stefano Laureti[3*]

[1]School of Engineering, University of Warwick
Library Road, Coventry, CV4 7AL, United Kingdom
e-mail:{S.Amato.1,D.A.hutchins}@warwick.ac.uk

[2]Centre for Offshore Engineering and Safety Technology, China University of Petroleum (East China), Qingdao, 266580, PR China
e-mail: xiaokang.yin@upc.edu.cn

[3]Department of Informatics, Modeling, Electronics and System Engineering, University of Calabria, Via P. Bucci, 87036 Arcavacata, Rende (CS), Italy
e-mail: {marco.ricci, stefano.laureti}@unical.it

*Abstract*— This paper introduces an improved image processing method usable in capacitive imaging applications. Standard capacitive imaging tends to prefer amplitude-based images over the use of phase due to better signal-to-noise ratios. The new approach exploits the best features of both types of information by combining them to form clearer images, hence improving both defect detection and characterization in non-destructive evaluation. The methodology is demonstrated and optimized using a benchmark sample. Additional experiments on glass fibre composite sample illustrate the advantages of the technique.

## I. INTRODUCTION

Capacitive imaging (CI) is an established non-destructive evaluation (NDE) technique for probing the integrity of different materials and structures [1,2]. CI has been demonstrated to be an effective contactless method for investigating reinforced concrete [3–5], dielectric composite materials such as ceramic armor tiles [6] and glass fibre reinforced polymer (GFRP) [7]. It can also be used to detect surface features on conductive composite materials such as carbon fibre reinforced polymer (CFRP) [7,8]. Moreover, CI has been successfully used to detect corrosion under insulation (CUI) [9,10]. Thus, CI spans a wide range of applications and can be used alongside more traditional NDE methods such as thermography, eddy-current, X-rays and ultrasonic testing [11].

The basic principle of the CI technique is the use of the electric fringing field from a pair of co-planar electrodes, created when an AC voltage is applied to the driving electrode. Any changes in the local electric properties within the volume probed will result in a change of the spatial distribution of the electric field and the induced charge at the sensing electrode. In the case of a dielectric material, a 2D image of the sensed volume is thus obtained by moving the electrodes across an area of the sample and then visualizing the amount of induced charge at selected locations.

The sensitivity distribution within the volume influenced by the fringing electric field has been found to be highly dependent on lift-off and the dielectric properties of the sample [12,13], as well as the probe/sensor geometry [14–18]. Moreover, the complex spatial spreading of the fringing electric field can cause changes in the 3D measurement sensitivity distribution with variations of the local properties of the sample, making the NDE of inhomogeneous and geometrically-complex material challenging. In addition, the sensitivity is found to change from positive values close to the CI probe to negative ones as the distance from the probe (i.e. depth within the sample) increases [19]. This unavoidable inversion makes the analysis of images difficult, as reported in the literature [8,12,13].

To overcome this issue, and to improve defect detection and characterization, the 3D measurement sensitivity distribution can be calculated by means of Finite Element Modeling, and its contribution to the CI's results/images taken into account by solving the inverse problem [20].

However, this approach is application-related and computationally expensive, as the sensitivity distribution must be calculated for each given sample and probe geometry. An alternative is the use of a planar array of capacitive imaging sensors, so that lift-off can be mitigated by inferring information from nearby electrodes [21], and/or the use of a lock-in amplifier to improve the Signal-to-Noise Ratio (SNR).

In this paper, an image processing and information fusion approach is introduced as a tool for improving both defect detection and characterization. This relies on the different information that can be inferred from both the amplitude and phase features, which are directly available when using a lock-in amplifier. It should be noted that images produced by CI are commonly obtained from plotting local amplitude values — images based on phase features have not received as much attention as they show lower SNR values [8,12] and the combination of the two does not seem to have been investigated.

We hereby show that amplitude and phase features can be fused together to form clearer images when compared to those obtainable using the two sets of information separately; further, defect detection is improved. Note that such techniques have been used in other NDE methods such as thermography [22], and the use of data fusion for NDE in general has also been described [23], where the choice of data fusion method was discussed in some detail. The general conclusion seemed to be that realistically several methods had to be tried for a given NDE method, and this is the approach described in this paper, where two different methods are explored.

The data fusion technique for CI is first tested on a Perspex sample containing defects with a known geometry and depth, as this can be used for both optimizing the technique and showing its working principle. It is shown that the proposed methodology yields an improved defect characterization in terms of depth estimation compared with standard amplitude-based images. The information fusion approach is then applied for the detection and evaluation of impact-damaged areas in three pultruded GFRP samples. The results are also compared with those obtained with air-coupled ultrasonic testing on the same sample and found to be in good agreement.

## II. Fundamentals of the CI technique

The CI technique relies on the indirect measure of the electric field spatial distribution variation inside the sample under inspection from a pair of co-planar electrodes to form a capacitor [5,7,9,12,19]. This is typically scanned across an area of the investigated sample to form an image. Although some imaging is performed at frequencies >1 MHz [11], an AC voltage within the 1 kHz -1 MHz frequency range is commonly applied to the driving electrode. Under this condition, Maxwell's equations describing the electromagnetic interaction among the co-planar electrodes and the surrounding media can be simplified by considering a quasi-static electric field and a negligible magnetic field. Thus, the local expression of Maxwell's equations for the quasi-static regime take the form of Eqs.(1-4):

$$\nabla \times \boldsymbol{E} \approx 0, \qquad (1)$$
$$\nabla \times \boldsymbol{H} \approx 0, \qquad (2)$$
$$\nabla \cdot \boldsymbol{D} \approx \rho, \qquad (3)$$
$$\nabla \cdot \boldsymbol{B} \approx 0. \qquad (4)$$

Thus, the electric field $\boldsymbol{E}$ of a charge distribution $\rho$ is irrotational, and the magnetic field magnitude $H$ is approximated by zero.

To understand how those relations are exploited in CI, Fig.1 shows a schematic of the interaction between the fringing electric field and a dielectric isotropic sample containing a defect. An AC voltage at a fixed frequency is applied to the driving electrode, producing a quasi-static electric field that penetrates into the sample. If the electrodes are shielded appropriately on their top surface, any changes in the electric field can be assumed to arise from the front surface of the electrodes in the form of closed loops. The local electric properties of the investigated materials induce a given amount of charge at the sensing electrode. Thus, the presence of a flaw within the probed volume changes the value of the local dielectric constant $\varepsilon$, and thus that of the induced charge at the sensing electrode. This amount is given according to both Gauss's law in Eq.(3) and to the constitutive relation $\boldsymbol{D} = \varepsilon\boldsymbol{E}$:

$$\nabla \cdot \boldsymbol{E} \approx \rho/\varepsilon, \tag{5}$$

meaning that scanning a pair of electrodes across an object and imaging the induced charge amplitude at the sensing electrode can be used to gather information about the probed volume.

## II. EXPERIMENTAL SETUP AND PROPOSED IMAGE FUSION APPROACH

Figure 2 depicts the experimental setup used in this work. A CI probe was scanned across the sample by means of a 2D motorized scanning stage, which moved it over a chosen grid of $N_x \times N_y$ points along the *x-y* directions in a horizontal plane. The lift-off in the *z* direction remained fixed throughout the scan. A function generator (Wavetek, model 191) was used to excite the driving electrode with a 20 V peak-to-peak sine wave signal at a specific frequency $f_{in}$, in this case chosen to be $f_{in}$=15 kHz. The same driving signal was sent to a digital lock-in amplifier (Stanford Research, model SR850), acting both as a reference waveform $V_{ref}(t)$ for the signal processing and as a trigger signal to synchronize the whole measurement chain. Note that the capacitance across the electrodes was not directly measured but was instead converted to an AC voltage via a charge amplifier (Cooknell, model CA/6C). The resultant signal was then input to a low-noise pre-amplifier (Stanford Research, model SR560) before being input into lock-in amplifier — this signal is hereinafter referred as $V_{out}(t)$.

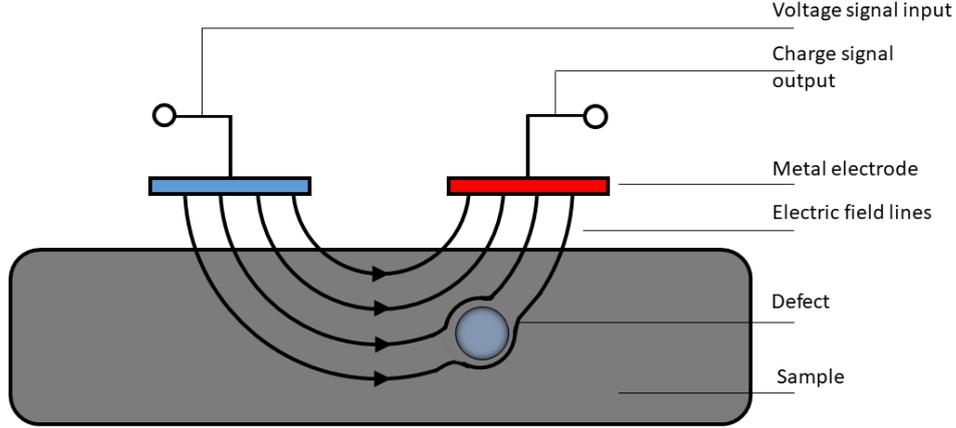

Fig. 1 A simplified sketch explaining the working principle of the CI technique: the fringing electric field is locally-perturbed by the presence of a defect, causing a measurable change in the induced charge at the sensing electrode.

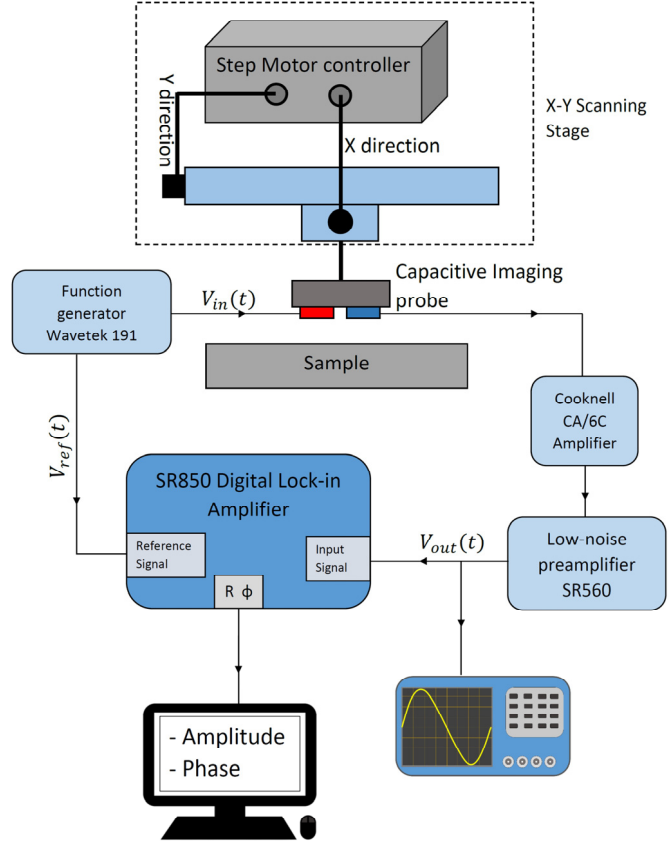

Fig. 2 Sketch of the experimental setup.

The signal $V_{out}(t)$ is in the form of a sinusoidal wave, together with additional noise $N(t)$, which is assumed to be spread over a wide frequency spectrum, *i.e.* much wider than theoretical single tone $f_{in}$:

$$V_{out}(t) = V_{out} \sin(\omega t + \theta_{out}) + N(t), \tag{6}$$

with $\theta_{out}$ being the phase of $V_{out}(t)$, and where $\omega = 2\pi f_{in}$. The lock-in amplifier is fed with the reference signal $V_{ref}(t) = V_{ref} \sin(\omega t + \theta_{ref})$, which is exploited to calculate both the in-phase $X$ and out-of-phase $Y$ contributions. As an example, the in-phase $X$ feature is obtained as follows:

$$\begin{aligned}V_{out}(t) \times V_{ref}(t) &= V_{out}V_{ref} \sin(\omega t + \theta_{out}) \sin(\omega t + \theta_{ref}) + N(t)\, V_{ref} \sin(\omega t + \theta_{ref}) = \\ &= {1}/{2}\, V_{out}V_{ref}[\cos(\theta_{out} - \theta_{ref}) - \cos(2\omega t + \theta_{out} - \theta_{ref})] + N(t)V_{ref} \sin(\omega t + \theta_{ref}).\end{aligned} \tag{7}$$

Thus, when using a low-pass filter, the in-phase feature (*X*) can be written as:

$$X = \frac{1}{2} V_{out} V_{ref} \cos(\theta_{out} - \theta_{ref}) \tag{8}$$

The corresponding out-of-phase contribution $Y$ can be calculated by shifting $V_{ref}$ by 90°, *i.e.* using $V'_{ref}(t) = V_{ref} \cos(\omega t + \theta_{ref})$ in Eq.(7), yielding:

$$Y = \frac{1}{2} V_{out} V_{ref} \sin(\theta_{out} - \theta_{ref}) \tag{9}$$

The amplitude $R$ and phase $\phi$ of the signal can be then obtained:

$$R = \sqrt{X^2 + Y^2}, \tag{10}$$

$$\phi = \tan^{-1}\left(\frac{Y}{X}\right) = \theta_{out} - \theta_{ref}, \tag{11}$$

Images can now be constructed by visualizing either $R(x,y)$ and $\phi(x,y)$ at each scanned location $(x,y)$. Note that only a few images have been reported using both the amplitude $R$ and phase $\phi$ [9,10], where the $\phi$ feature was simply discarded as it shows usually lower *SNR* level than $R$-based images. Also, in [9], the authors stated that the identification of some defects in several benchmark samples was only possible thanks to *a priori* knowledge of their location. Moreover, most of the CI results reported in the literature rely on the proper selection of threshold values for the recorded amplitude of $R$, and in addition abrupt changes are often seen due to edges effect, lift-off variations and geometrical complexities. These effects are unavoidable when scanning real samples in in-situ applications.

A procedure is thus introduced which relies on the fusion of $R$ and $\phi$ to enhance both defect identification and the selection of a proper threshold for imaging, as well as a more accurate visualization of the features of interest. This is achieved by carried out the following steps:

1) Normalising both $R$ and $\phi$ matrixes to [0,1] by min-max normalisation.

2) Fusing $R_{norm}$ and $\phi_{norm}$ in the two following ways:

$$\Delta = \phi_{norm} \times (1 - R_{norm}), \tag{12}$$

$$\Xi = \phi_{norm} R_{norm}^{-1}. \tag{13}$$

It will be shown that a unique feature detection, *e.g.* defect localization, can be achieved by imaging using $\Xi$, whilst a better defect evaluation can be achieved by imaging after calculating $\Delta$. It is worth noting that different amplification chains could led to a sign change in the phase feature. In this case, the same results here reported could be obtained by defining $\Delta' = (1 - \phi_{norm}) \times (1 - R_{norm})$, and $\Xi' = (1 - \phi_{norm}) R_{norm}^{-1}$.

III. DESCRIPTION OF THE INVESTIGATED SAMPLES AND CI PROBES

A combination of samples having different types of defects and probes geometries have been used so as to show the robustness and capabilities of the proposed image fusion approach in the most general sense. The first is a Perspex sample containing defects in the form of flat-bottomed holes, which has been used as a benchmark to test the technique thoroughly. Figure 3 depicts the dimensions of this specimen:

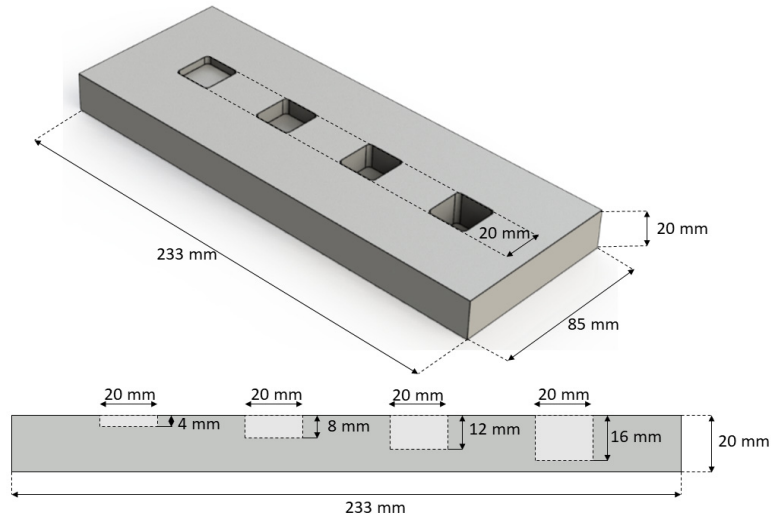
Fig. 3 Perspex sample geometry and dimension.

In addition, three EXTREN® (series 500) pultruded GFRP samples (with fiberglass reinforcement and thermosetting polyester) were available containing various levels of impact damage caused by 14 J, 16 J, and 18 J impact energies from an impactor with a hemispherical tip. The surfaces of the three samples are shown in Figs. 4(a), (b) and (c), respectively. Note that the impact-damaged areas are marked with black crosses on these images, as they are barely visible. For each sample, images were also obtained by air-coupled ultrasonic testing [24] to indicate the dimension and shape of the impacted area. These will be used as a comparison for evaluating the capabilities in defects sizing of the proposed imaging procedure.

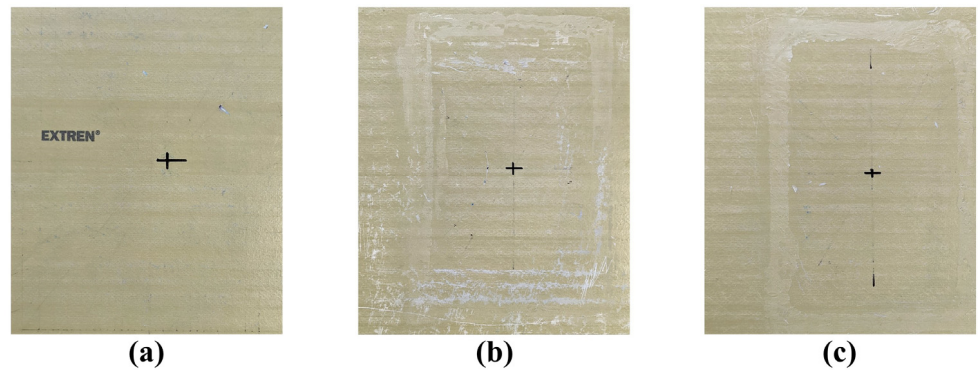

(a)   (b)   (c)

Fig. 4 Pultruded GFRP samples with damage caused by (a) 14 J, (b) 16 J, and (c) 18 J impact energies at their centres.

The samples shown in Figs. 3-4 have been investigated using two different CI probe geometries. These are a back to back electrode arrangement (Fig. 5(a)) and one with a concentric geometry (Fig. 5(b)). For the back-to-back probe in Fig.5(a), the parameters *s*, *b* and *h* were equal in value to 4, 16 and 19 mm, respectively, while R1, R2 and R3 where equal to 8, 16, 24 mm for the concentric one. In this case, the inner electrode was set to be the source. Note that it was thought interesting to compare these two commonly-used geometries, even though the concentric probe would be likely to produce lower-resolution images due to its size and the fact that the electric field distribution is not contained principally along one axis but is created radially from the central source electrode to the outer annulus.

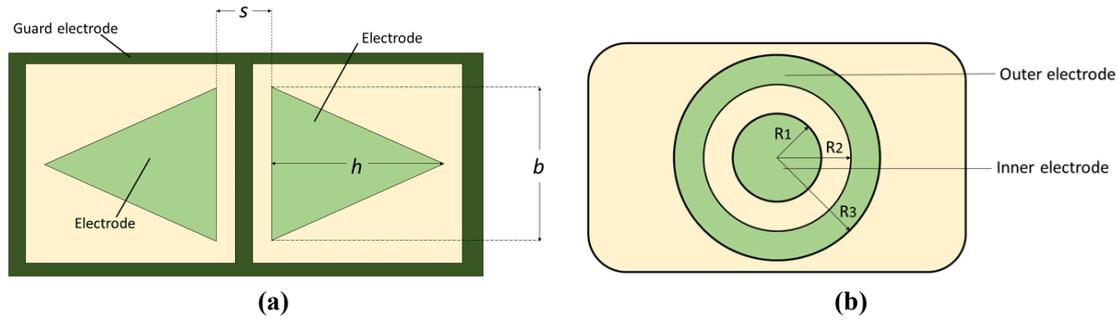

(a)                                           (b)

Fig. 5 The two CI probes geometries: (a) the back-to back probe, and (b) the concentric geometry.

## IV. RESULTS

### A. The Perspex sample

Figures 6 (a,c) show the CI results obtained by imaging $R$ and $\phi$ with the back-to-back probe (Fig. 5(a)) with a vertical lift-off distance $z$ of 3 mm. Note that the probe main axis was always parallel to the $x$-axis, as shown in the figure. Figures 6(a,c) are a scan of the whole sample, whereas Figs. 6(b,d) show the improved images that were obtained by both limiting the results to areas far from the sample edges and by choosing a correct thresholding value for the image.

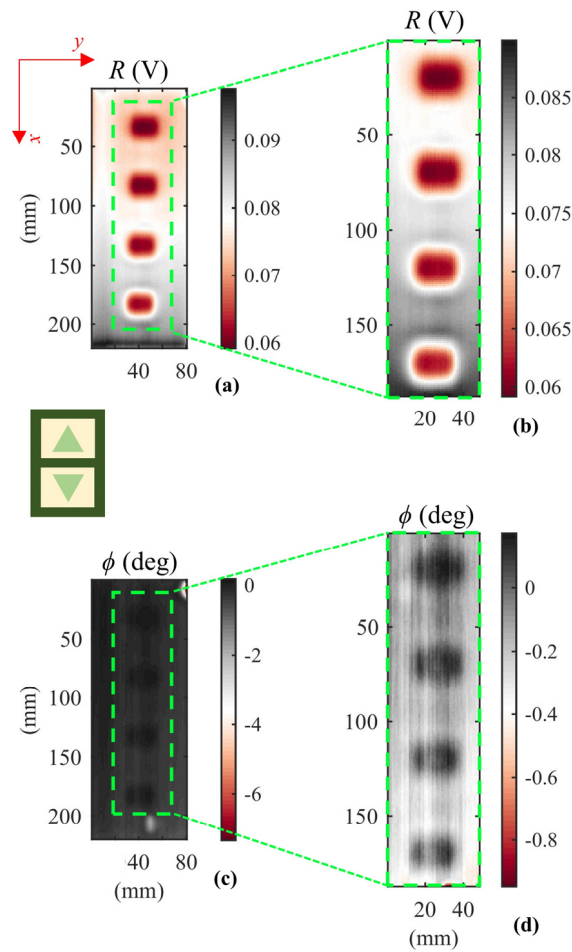

Fig. 6 Capacitive images for the Perspex sample shown earlier in Fig. 3 using the back-to-back electrodes of Fig. 5(a). (a) Amplitude ($R$) image; (b) $R$ image for a smaller region away from the sample edges; (c) Phase-based ($\phi$) image; (d) $\phi$ image for a smaller region away from the sample edges.

Although improvements in imaging are possible by careful selection of scan area and thresholding values, a rapid inspection and image analysis is often highly desirable when evaluating a real sample over its whole area, especially in an industrial environment. As shown in Fig. 7, this is possible using the proposed data fusion approach by making use of both $R$ and $\phi$ values. Here, a series of images are presented, obtained using both electrode types shown earlier in Fig. 6, and by imaging the whole sample without any area selection to remove sample edge effects. It can be seen that the proposed fusion approach to generate experimental values for both $\Xi$ (eqn. (12)) and $\Delta$ (eqn. (13) can lead to enhanced imaging. The properties of the $\Xi$ variable allow all the Perspex defects to be identified by both the back-to-back probe and circular probes. The $R$ images from the back-to-back probe (Fig. 7(a)) have a higher resolution than that from the concentric probe (Fig. 7(b)), as expected from the argument above. Note also that the phase information $\phi$ is much better for the back-to-back probe, the concentric probe tending to even out phase variations. In both cases, it can be seen that the $\phi$ images have lower signal to noise values than those for $R$, as has been observed by other authors [8,12]; however, the data fusion results leading to images based on $\Xi$ and $\Delta$ produce improvements over the more conventional $R$ and $\phi$ images, as expected. In particular, the defect is more easily detected in the concentric electrode case, even though the phase image does not enable the defect to be seen clearly. The lack of visible phase information for the concentric probe is understandable — in this case the electric field is radial, and the phase information would be an overall result over a 360º angular range. Conversely, the back-to-back design concentrates the field in one direction only, producing better phase information.

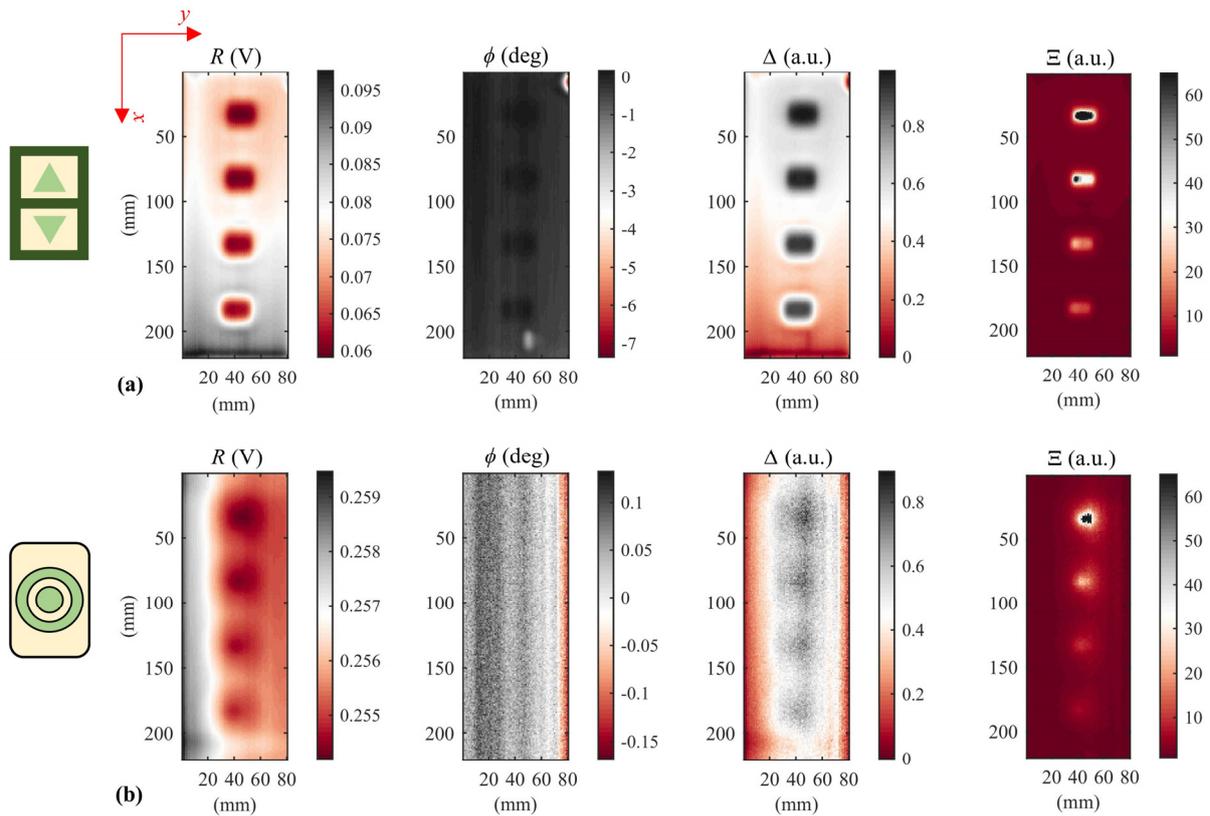

Fig. 7 Experimental results for scans on the Perspex sample using (a) the back-to back probe, and (b) the concentric design.

It is interesting to study the contribution of $\Delta$ to the defect characterization. To illustrate this, the normalized values of $R$ and $\Delta$ along the same line of pixels crossing all the defects are compared in Figure 8 (a). At first glance, both the trends are very similar. However, Perspex defects have a depth values increasing linearly from the shallowest one (Number 4) to the deepest one (Number 1), see Fig. 3. Therefore, obtaining a feature showing a linear relation with those defects is of interest in that it illustrates the effectiveness of the imaging procedure. Further, Fig. 8(b) thus compares the normalized amplitude peak values for both $R$ and $\Delta$ at the location of maximum sensitivity to a given defect. The linear fits and their root mean square errors (RMSE) values, quoted on the figure, demonstrate that Δ is better than the standard $R$-based estimation, for which a higher order polynomial fitting would be more accurate. This means that the phase information is perhaps more beneficial for inferring defects features such as depths. Finally, note that the images obtained with the concentric electrodes show poor SNR values — this is due to the fact the electrodes dimension are bigger than that of the defect. It can be also noted that, due to the geometry of the electrodes, the phase is less sensitive because the electrostatic field is averaged over a larger area.

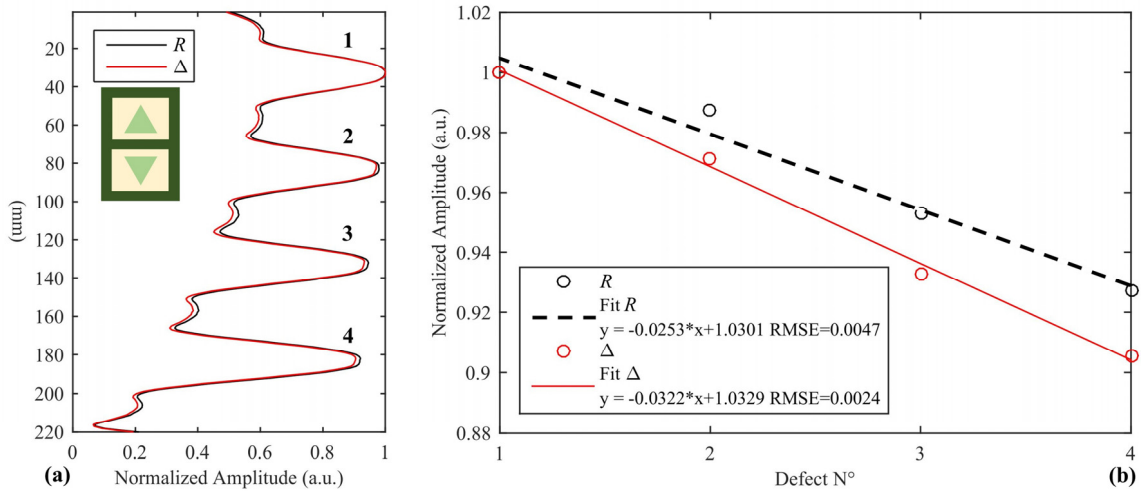

Fig. 8 (a) Normalized amplitude of $R$ and Δ along a line crossing the centre of the defects (b) maximum values of $R$ and Δ for each defect to demonstrate linearity.

## B. The impact-damaged GFRP samples

Figures 9 (a)-(c) show CI results for $R$ and $\phi$ obtained for the three GFRP samples shown earlier in Figure 4, with defects caused by impact with 14 J, 16 J and 18 J energies respectively. These were taken with the back-to-back probe design. It is clear that the defects have been detected in each case, especially by imaging the $R$ feature.

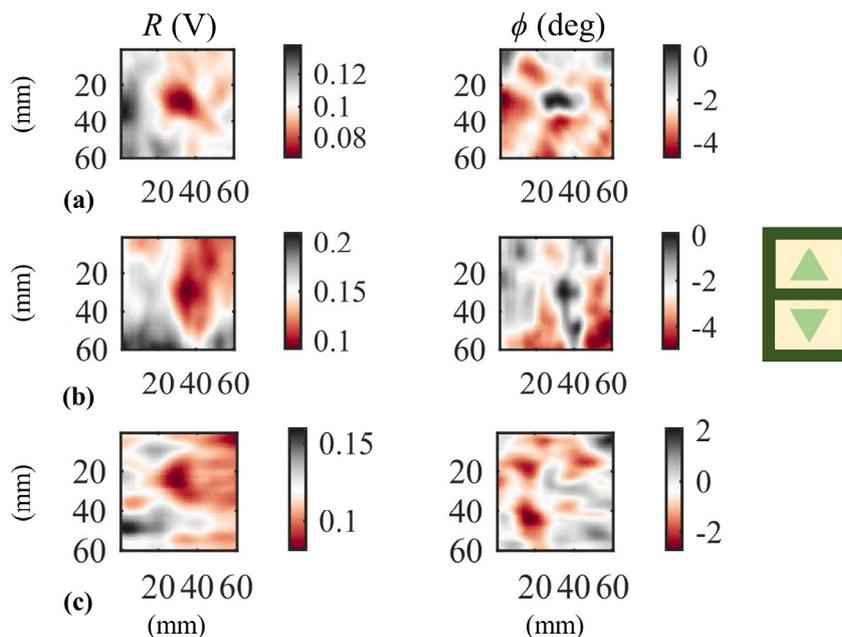

Fig. 9 Experimental scan results for $R$ and $\phi$ for the three samples with impact damage caused by impact energies of (a) 14 J, (b) 16 J and (c) 18 J.

The result of fusing data to obtain Δ and Ξ for the damaged area is shown in Figures 10 (a)-(c), where now the defects are more clearly visible in both sets of images. Also shown in Figure 10 is a comparison to a separate experiment where the defects were images using an air-coupled ultrasonic (ACU) system in through-transmission [23]. This used 10 mm diameter ultrasonic transducers, and hence would be expected to result in an image where the defect is visible over a larger area. It can be seen that the Δ and Ξ images are as expected of smaller dimension, with the Ξ image in particular giving a much better idea of the location of the defect, and the Δ image its size, than was available from the raw amplitude ($R$) and phase ($\phi$) data before the fusion process.

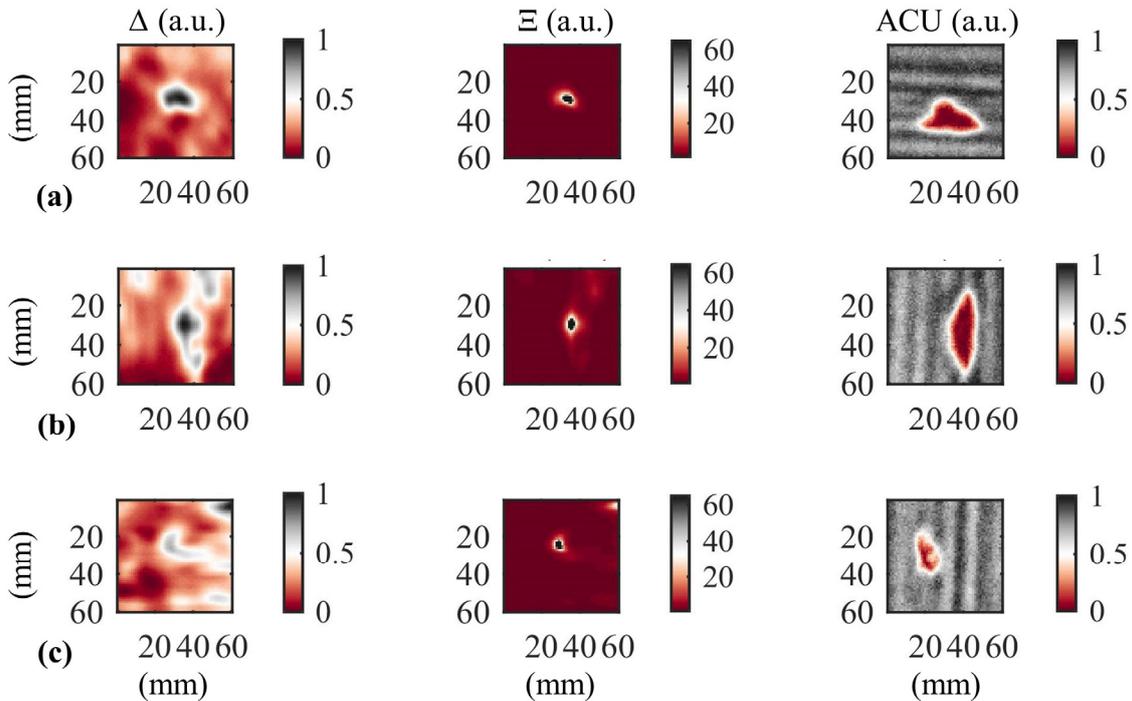

Fig. 10: Experimental scan results for Δ and Ξ for the three samples with impact damage caused by impact energies of (a) 14 J, (b)16 J and (c) 18 J. Also shown is the result of an air-coupled ultrasound (ACU) scan of the same samples for comparison.

Figure 11 shows the results obtained using the concentric probe design on the 14 J GFRP impact damaged sample. Again, the poor SNR achieved with the phase ($\phi$) image was due to the dimensions and radial geometry of the probe, although this became somewhat more visible in the Δ image after data fusion. However, the Ξ image, as in the previous cases described here, demonstrated a very good retrieval of the image location. This is interesting, as it would allow the larger concentric probe to be used for rapid scanning of large objects without having to consider alignment of the probe geometry (as would be needed for the back-to-back design). More detailed scanning via the back-to-back probe could then get a better idea of its lateral dimensions.

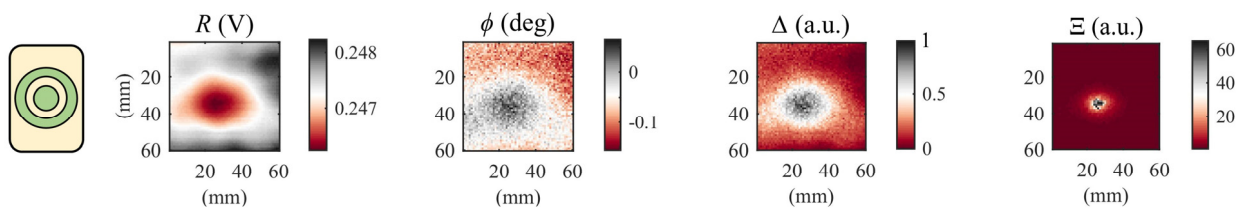

Fig. 11 CI with concentric probe on 14 J.

V. CONCLUSIONS

This study has introduced an improved method for imaging defects in CI for NDE. In general, conventional CI imaging tends to use amplitude over phase information to form the images, as the phase-based images show a higher noise level. The present method is based on fusing these two features in order to mitigate events such as edge effects, lift-off variations and geometrical complexities, which may affect resultant images hiding features of interest. Preliminary tests aimed to demonstrate and optimize the image processing have been conducted on a Perspex sample with flat-bottomed holes. The findings show that fusing amplitude and phase information provides both a unique defect localization and a better defect evaluation. Further tests on composite GFRP samples containing impact damage have demonstrated that the technique is very useful in allowing the two probe designs to be used — the concentric probe for rapid defect detection and location, and the back-to-back probe for more detailed defect characterisation. The results on GFRP samples were compared with images obtained from a pulse-compression air coupled ultrasonic measurement, illustrating the robust nature of the imaging approach. It is thought that this use of data fusion allows capacitive imaging to be used on a wider range of composite samples and defects.

Acknowledgements

This work is funded by the European Union's Horizon 2020 Research and Innovation programme under the Marie Skłodowska-Curie grant agreement No 722134 – NDTonAIR.